# Chapter 2
# UV filaments

*Ali Rastegari,[1,2] Alejandro Aceves,[3] and Jean-Claude Diels[1,2]*

## Abstract

This chapter starts with a discussion of the main qualitative differences between UV and mid-IR filaments: from multiphoton ionization in the UV to tunnel ionization in the near- to mid-IR. A general qualitative analysis of the properties of single filaments versus wavelength follows. Because of their long pulse duration, a quasi-steady-state theory of their propagation is possible. An eigenvalue approach leads to a steady-state field envelope that is compared to the Townes soliton. However, that solution is close enough to a Gaussian shape to justify a parametric evolution approach. After this theoretical introduction, an experimental verification at 266 nm follows. Femtosecond UV filaments were generated with frequency-tripled Ti:sapphire sources and KrF amplifiers. The source for long-pulse filaments is an oscillator-amplifier Nd-YAG Q-switched system, frequency doubled, compressed, and frequency doubled again to reach 170 ps pulses of 300 mJ energy. The sub-nanosecond duration of the UV pulse may revive the debate as whether the filament is a moving focus or self-induced waveguide. Two applications of UV filaments are presented in the last two sections. It is shown that isotopically selective laser-induced breakdown spectroscopy (LIBS) is possible by exploiting the narrow dips observed in the emission spectrum. These dips are due to reabsorption by the material in the plume created by the impact of the filament on a solid. These absorption lines are only a few-pm wide and are exactly centered at the wavelength of a transition from ground state of the material, without any Stark shift or broadening. A final application is laser-induced discharge, which is a guided discharge that follows exactly the path of the inducing UV filament. Laser-induced discharge may lead to the control of lightning, which is a topic of intense research in Europe.

[1]Department of Physics, The University of New Mexico, Albuquerque, USA
[2]Center for High Technology Materials, Albuquerque, USA
[3]Southern Methodist University, Dallas, USA

## Acknowledgments

This work is supported by the Army Research office (ARO: W911NF-19-1-0272) and US Department of Energy (DOE: DESC0011446).